\documentclass[twocolumn,preprintnumbers,amsmath,amssymb]{revtex4}
\usepackage{amsfonts}
\usepackage{graphicx}

\begin{document}

\title{Cut-wire pairs and plate pairs as magnetic atoms for optical metamaterials}

\author{G. Dolling, C. Enkrich, and M. Wegener}
\affiliation{Institut f\"ur Angewandte Physik, Universit\"at Karlsruhe (TH), Wolfgang-Gaede-Stra{\ss}e 1, D-76131 Karlsruhe, Germany}
\author{J. Zhou and C.\,M. Soukoulis$^*$}
\affiliation{Ames Laboratory and Department of Physics and Astronomy, Iowa State University, Ames, Iowa 50011, U.S.A.}
\author{S. Linden}
\affiliation{Institut f\"ur Nanotechnologie, Forschungszentrum Karlsruhe in der Helmholtz-Gemeinschaft, Postfach 3640, D-76021 Karlsruhe, Germany}

\begin{abstract}
We study the optical properties of metamaterials made from
cut-wire pairs or plate pairs. We obtain a more pronounced optical
response for arrays of plate pairs -- a geometry which also
eliminates the undesired polarization anisotropy of the cut-wire
pairs. The measured optical spectra agree with simulations,
revealing negative magnetic permeability in the range of
telecommunications wavelengths. Thus, nanoscopic plate pairs might
serve as an alternative to the established split-ring resonator
design.
\\[5pt]
\copyright  2005 Optical Society of America
\end{abstract}


\maketitle
Magnetic and left-handed metametarials have recently attracted considerable attention \cite{Shelby2001,Yen2004,Smith2004,Linden2004,Zhang2005,Moser2005,Crete2005}
because of their unusual optical properties \cite{Veselago1968} and because of potential applications, such as, for example, "perfect lenses".\cite{Pendry2000}
The latter require a negative refractive index, which can be realized by a negative electric permittivity and a negative magnetic permeability
-- at the same frequency. Traditionally,\cite{Pendry1999} this is achieved by a combination of artificial ``electric atoms'' and artificial ``magnetic atoms'',
i.e., by split-ring resonators and metallic wires, with ``lattice constants'' much smaller than the wavelength of light, such that the light
field experiences an effective homogeneous medium.

Other theoretical work \cite{wirepair1,wirepair2,wirepair3} showed, however, that pairs of finite-length wires (``cut-wire pairs'')
would not only be able to replace the split-ring resonators (SRR)
but would possibly also lead to a negative refractive index directly -- without the need for additional metallic wires.
It is the aim of this letter to investigate the optical properties of metamaterials made from such cut-wire pairs and from
related plate pairs experimentally and
to systematically study the dependence on different design parameters.

\begin{figure}[h]
\centerline{\scalebox{0.75}{\includegraphics[width=8.2cm,keepaspectratio]{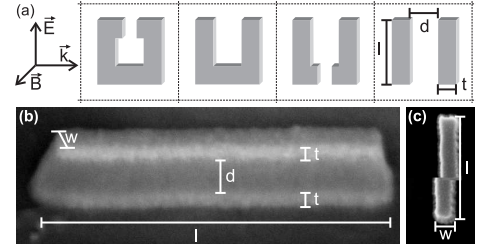}}}
\caption{(a) Scheme illustrating the ``adiabatic'' transition from
split-ring resonators (LHS) to cut-wire pairs (RHS) as ``magnetic
atoms'' of optical metamaterials. (b) Electron micrograph
(oblique-incidence view) of an actual cut-wire pair with
$w=150\,\rm nm$, $t=20\,\rm nm$, $d=60\,\rm nm$, $l=700\,\rm nm$,
(c) corresponding top view.}
\end{figure}

Fig.\,1 (a) illustrates the connection between usual SRR and cut-wire pairs. On the LHS, a usual SRR is shown, which can be viewed as just one winding of a
magnetic coil with inductance $L$ in series with a capacitance $C$.\cite{Linden2004} The latter is formed by the ends of the wire. The oscillating magnetic
field perpendicular to the SRR plane induces a circulating current in the coil. Near to the $LC$ resonance at angular frequency $\omega_{LC}=1/\sqrt{LC}$,
the circulating current in the coil can lead to a magnetic moment perpendicular to the plane of the coil that can counteract
the external magnetic field, enabling a negative magnetic permeability. Opening up the split in the SRR decreases the capacitance $C$,
hence increases the $LC$ resonance frequency. Additionally opening the bottom arm of the resulting ``U'' leads to a second serial capacitance,
further reducing the net capacitance in the circuit. This results in an increase of the resonance frequency.
Further opening of the lower slit brings us to a pair of cut wires. The result of this transition is that the Ohmic currents
in the horizontal arms on the LHS of Fig.\,1 (a) have been replaced by displacement currents on the RHS.
On the one hand, we have increased the $LC$ resonance frequency for a given minimum feature size. On the other hand, this
increased resonance frequency at fixed lattice constant decreases the ratio between (resonance) wavelength, $\lambda$, and lattice constant, $a$.
A true metamaterial requires a very large ratio of wavelength to lattice constant, typically on the order
of $\lambda/a=10$ for SRR.\cite{retrieval_Smith,retrieval_Koschny} With the cut-wire-pair design, this ratio is typically $\lambda/a \approx 2$.

Obvious relevant design parameters are the length of the cut wires $l$, the metal thickness $t$, the width $w$ of the wires, and the spacing between the wires $d$.
To fabricate structures suitable for the polarization configuration depicted in Fig.\,1(a) on a substrate for normal incidence conditions,
the cut wires have to be on top of each other, with a certain dielectric spacer layer in between (see Fig.\,1(b)). Such design is straightforward to realize:
We start by defining the lateral structure in a PMMA photoresist (spin-coated onto a glass substrate covered with 5-nm layer of indium-tin-oxide)
using standard electron-beam lithography.
Thereafter, we sequentially deposit a $t=20$-nm thin gold layer, a $\rm MgF_2$ layer of thickness $d$ and refractive index 1.39,
and another 20-nm thin gold layer, all via electron-beam evaporation, followed by a lift-off procedure.
All samples presented in this letter have a total area of $(80\,{\rm \mu m})^2$.

Transmittance and reflectance spectra are measured with a Fourier-transform infrared spectrometer
(Bruker Equinox 55, NIR halogen source) combined with an infrared microscope (Bruker Hyperion 1000, $36 \times$ cassegrain lens, numerical aperture
${\rm NA} = 0.5$, Si detector or liquid $N_2$-cooled InSb detector, infrared polarizer). The samples are aligned with their surfaces perpendicular
to the optical axis. The transmittance and reflectance spectra are normalized to the bare substrate and a silver mirror,
respectively, and are taken for two orthogonal linear polarizations of the incident light.

\begin{figure}[h]
\centerline{\scalebox{0.85}{\includegraphics[width=8.2cm,keepaspectratio]{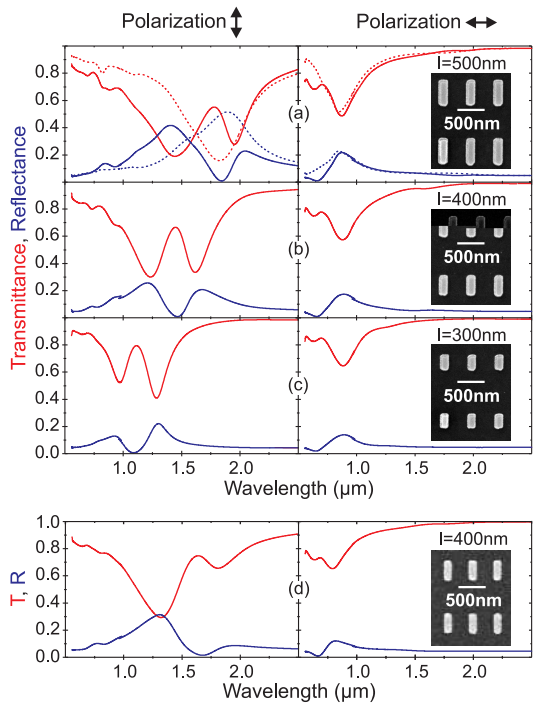}}}
\caption{Measured transmittance (red) and reflectance (blue)
spectra for cut-wire pairs for vertical incident polarization (LHS
column) and horizontal polarization (RHS column). Parameters
varied: (a) $l=500\,\rm nm$, (b) $l=400\,\rm nm$, and (c)
$l=300\,\rm nm$. Fixed parameters for (a)--(c): $w=150\,\rm nm$,
$t=20\,\rm nm$, $d=80\,\rm nm$, $a_x=500\,\rm nm$, and
$a_y=1050\,\rm nm$. The dotted curves in (a) are spectra from a
nominally identical structure, but without the upmost gold wire.
(d) as (b), but $d=60\,\rm nm$ rather than 80\,nm. The insets in
(a)-(d) show corresponding electron micrographs (top view).}
\end{figure}

Fig.\,2 (a)-(c) show results for cut-wire pairs of different length but fixed $\rm MgF_2$ spacer thickness and fixed wire width.
The dotted curves in (a) correspond to a nominally identical sample, however, {\it without} the upmost gold layer.
This single cut-wire sample shows only {\it one} pronounced resonance -- the Mie resonance -- for each polarization.
For an incident polarization along the long axis of the cut-wire pairs, {\it two} resonances are observed,
which essentially disappear for orthogonal polarization.
If the two wires were identical, had identical environments, and were excited equivalently,
these two resonances would correspond to the symmetric (antisymmetric) low-frequency
(high-frequency) mode of the coupled system of the two cut wires. However, in our case the symmetry is already broken by the
excitation geometry and by the presence of the substrate. As a result,
both resulting resonances have antisymmetric character to some extend, thus, both of them have
a corresponding magnetic dipole moment connected with a resonance in the magnetic permeability $\mu$.
A comparison of Fig.\,2 (b) and (d) shows the dependence on the $\rm MgF_2$ spacer thickness $d$.
As expected from the above picture of two coupled oscillators, the splitting between the two effective resonances depends on their coupling:
For thin (thick) spacers, the coupling is strong (weaker), hence the two resonances are split by a large (smaller) amount in the spectrum.

\begin{figure}[t]
\centerline{\scalebox{0.85}{\includegraphics[width=8.2cm,keepaspectratio]{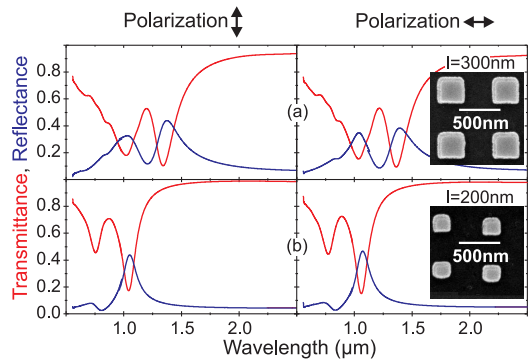}}}
\caption{Measured transmittance (red) and reflectance (blue)
spectra for arrays of plate pairs. Parameters varied: (a)
$w=l=300\,\rm nm$ and (b) $w=l=200\,\rm nm$. Fixed parameters:
$t=20\,\rm nm$, $d=80\,\rm nm$, and $a_x=a_y=l+350\,\rm nm$.
Representation as in Fig.\,2.}
\end{figure}

The obvious polarization dependence of the cut-wire pairs is undesired, for example in potential applications as ``perfect lenses''.
Thus we have also investigated samples for which the wire width equals the wire length, i.e., where $w=l$. In this case, the cut-wire pairs turn into nanoscopic
plate pairs. Their measured optical properties are shown in Fig.\,3. Obviously, the resulting optical resonances are qualitatively similar, yet
even more pronounced than in the case of cut-wire pairs. Reducing the value of $w=l$, allows for tuning of the resulting resonance positions.

\begin{figure}[t]
\centerline{\scalebox{0.85}{\includegraphics[width=8.2cm,keepaspectratio]{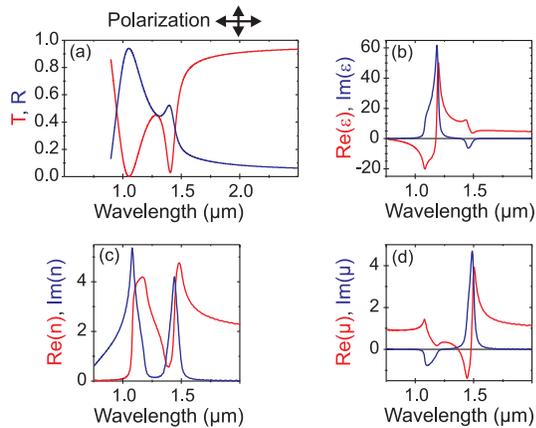}}}
\caption{(a) Calculated transmittance $T$ (red) and reflectance
$R$ (blue) spectra for the structure shown in Fig.\,3(a), i.e.,
for a single layer. Vertical and horizontal incident linear
polarization are strictly identical. (b) Electric permittivity
$\epsilon$, (c) refractive index $n$, and (d) magnetic
permeability $\mu$ as obtained from the corresponding retrieval
for a stack of layers.}
\end{figure}

\newpage

For the mentioned reasons, the samples corresponding to Fig.\,3(a) seem most attractive to us.
Thus, we explicitly present the retrieval of the magnetic permeability $\mu$,
the electric permittivity $\epsilon$, and the refractive index $n$ here. We use the Drude model for the gold \cite{Linden2004} and all other parameters
as described above.
The retrieval is done along the traditional lines,\cite{retrieval_Smith,retrieval_Koschny} i.e.,
we calculate transmittance and reflectance spectra for a sequence of layers perpendicular to the substrate, i.e.,
for a periodic medium with lattice constant $a_z=130\,\rm nm$ and retrieve $\mu$, $\epsilon$, and $n$ from that.
Fig.\,4 shows the calculated transmittance and reflectance spectra as well as the retrieval for the parameters of the structure shown in Fig.\,3(a).
For the long-wavelength resonance, we obtain a spectral region with a negative real part of $\mu$ indeed,
but no negative ${\rm Re}(\epsilon)$ at the same frequency and no negative ${\rm Re}(n)$ (red curves). The corresponding imaginary parts
of $\mu$, $\epsilon$, and $n$ (blue curves) are also depicted. The retrieval for the short-wavelength resonance and beyond
should be taken with some caution as one gradually leaves the metamaterial limit, i.e., periodicity effects come into play.

In conclusion, we have experimentally realized and characterized nanoscale cut-wire pairs and plate pairs,
both of which are alternatives for split-ring resonators as ``magnetic atoms'' in metamaterials.
Plate pairs are advantageous compared to cut-wire pairs because their optical properties are identical for the two orthogonal
incident linear polarizations. Comparison with theory and subsequent retrieval shows that both options exhibit a frequency range
with negative magnetic permeability. In contrast to other reports,\cite{otherclaims}  however, we do not obtain a
negative refractive index from the cut-wire pairs or plate pairs alone, that we have studied so far.

We acknowledge support by the Deutsche Forschungsgemeinschaft
through subproject A1.5 of the DFG-Forschungszentrum ``Functional
Nanostructures'' (CFN) and by project We\,1497/9-1. The research of C.\,M.\,S. is further supported by the
Alexander von Humboldt senior-scientist award 2002, by Ames Laboratory (Contract No. W-7405-Eng-82),
EU FET project DALHM and DARPA (HR0011-05-C-0068).

$^*$ C.M.S. is also at Institute of Electronic Structure and Laser at FORTH, Heraklion, Crete, Greece.



\begin{thebibliography}{99}
\bibitem{Shelby2001}  R.\,A. Shelby, D.\,R. Smith, and S. Schultz, Science {\bf 292}, 77 (2001).
\bibitem{Yen2004} T.\,J. Yen, W.\,J. Padilla, N. Fang, D.\,C. Vier, D.\,R. Smith, J.\,B. Pendry, D.\,N. Basov, and X.\,Zhang, Science {\bf 303}, 1494 (2004).
\bibitem{Smith2004} D.\,R. Smith, J.\,B. Pendry, and M.\,C.\,K. Wiltshire, Science {\bf 305}, 788 (2004).
\bibitem{Linden2004} S. Linden, C. Enkrich, M. Wegener, J. Zhou, T. Koschny, and C.\,M. Soukoulis, Science {\bf 306}, 1351 (2004).
\bibitem{Zhang2005} S. Zhang, W. Fan, B.\,K. Minhas, A. Frauenglass, K.\,J. Malloy, and S.\,R.\,J. Brueck, Phys. Rev. Lett. {\bf 94}, 37402 (2005).
\bibitem{Moser2005} H.\,O. Moser, B.\,D.\,F. Casse, O. Wilhelmi, and B.\,T. Saw, Phys. Rev. Lett. {\bf 94}, 63901 (2005).
\bibitem{Crete2005} N. Katsarakis, G. Konstantinidis, A. Kostopoulos, R.\,S. Penciu, T.\,F. Gundogdu, M. Kafesaki, E.\,N. Economou,
Th. Koschny, and C.\,M. Soukoulis, Opt. Lett. {\bf 30}, 1348 (2005).
\bibitem{Veselago1968} V.\,G. Veselago, Sov. Phys. Usp. {\bf 10}, 509 (1968).
\bibitem{Pendry2000}  J.\,B. Pendry, Phys. Rev. Lett. {\bf 85}, 3966 (2000).
\bibitem{Pendry1999}  J.\,B. Pendry, A.\,J. Holden, D.\,J. Robbins, and W.\,J. Stewart, IEEE Trans. Microwave Theory Tech. {\bf 47}, 2075 (1999).
\bibitem{wirepair1} A.\,N. Lagarkov and A.\,K. Sarychev, Phys. Rev B {\bf 53}, 6318 (1996).
\bibitem{wirepair2} L.\,V. Pamina, A.\,N. Grigorenko, and D.\,P. Makhnovskiy, Phys. Rev B {\bf 66}, 155411 (2002).
\bibitem{wirepair3} V.\,A. Podolskiy, A.\,K. Sarychev, E.\,E. Narimanov, and V.\,M. Shalaev, J. Opt. A: Pure Appl. Opt. {\bf 7}, 32 (2005).
\bibitem{retrieval_Smith} D.\,R. Smith, S. Schultz, P. Markos, and C.\,M. Soukoulis, Phys. Rev. B {\bf 65}, 195104 (2002).
\bibitem{retrieval_Koschny} Th. Koschny, P. Markos, E.\,N. Economou, D.\,R. Smith, D.\,C. Vier, and C.\,M. Soukoulis, Phys. Rev. B {\bf 71}, in press (2005).
\bibitem{otherclaims} While completing this manuscript we learned about independent and related work on cut-wire pairs:
V.\,M. Shalaev, International Conference on Quantum Electronics and Laser Science (QELS), Baltimore (U.S.A.), May 22-27, 2005, JThA1.
\end{thebibliography}
\end{document}